\documentclass[12pt]{article}

\usepackage{geometry}
\usepackage{booktabs}
\usepackage{longtable,lscape}
\usepackage{graphicx}
\usepackage{multicol}
\usepackage{setspace}
\usepackage[UKenglish]{babel}
\usepackage{natbib}
\usepackage{tabularx}

\def\b#1{\mbox{\boldmath $#1$}}
\def\bl#1{\mbox{\footnotesize \boldmath {$#1$}}} 

\begin{document}
\title{\vspace*{-2cm}
An investigation  of the discriminant power and dimensionality of
items used for assessing health condition of elderly people}
\author{F. Bartolucci\footnote{Department of economics, Finance and Statistics, University of
Perugia, IT} \footnote{email: bart@stat.unipg.it}, G.
d'Agostino$^*$\footnote{email: gdago@stat.unipg.it}, G.E.
Montanari$^*$\footnote{email: giorgio@stat.unipg.it}}

\maketitle
\begin{abstract}
\begin{singlespace}
With reference to the questionnaire adopted within the Italian
project ``Ulisse" to assess health condition of elderly people, we
investigate two important issues: discriminant power and actual
number of dimensions measured by the items composing the
questionnaire. The adopted statistical approach is based on the
joint use of the latent class model and a multidimensional item
response theory model based on the 2PL parametrization. The latter
allows us to account for the different discriminant power of these
items. The analysis is based on the data collected on a sample of
1699 elderly people hosted in 37 nursing homes in Italy. This
analysis shows that the selected items indeed measure a different
number of dimensions of the health status and that they considerably
differ in terms of discriminant power (effectiveness in measuring
the actual health status). Implications for the assessment of the
performance of nursing homes from a policy-maker prospective are
discussed.\vspace*{0.25cm}
\end{singlespace}
\end{abstract}

\newpage


\section{Introduction}

The progressive ageing of the contemporary society, due to the
increasing life expectation, has raised the demand for health
assistance, stimulating the debate on the quality of the care
provided by nursing homes. According to the OECD, the percentage of
elderly people in the population of the industrialized countries
will increase from $13\%$ of 2000 to $25\%$ of 2040. As a
consequence, the cost of health care, especially for long term care
assistance, will rapidly increase. \cite{Anderson2000}, considering
the eight most industrialized OECD countries, found that the cost
for care assistance of elderly people increased from $3\%$ to $5\%$
of the GDP between 1993 and 1999 and the cost for long term health
assistance increased from $0.9\%$ and $1.6\%$ in the same period. In
Italy, public spending for long term assistance in 2002 was, in
terms of GDP, about $0.7\%$, which is expected to strongly increase
in the next decade.

A direct consequence of the increasing cost for health assistance is
that the Governments of industrialized countries have begun to
consider the problem of the rationalization of public interventions,
in terms of public spending, regulation, and policies. Public
intervention is also crucial to guarantee the accessibility to care
facilities of elderly people with low income, without affecting the
economic status of their families. Since the access to long term
care assistance is conditioned on the possibility to pay for the
services provided by nursing homes, there could be an artificial low
level of demand for long term care assistance; see
\cite{Alecxih1994}.

The above arguments imply that methodological tools to analyze the
performance of facilities for care assistance of elderly people are
of central interest for policy-makers. The aim is twofold: ({\em i})
to promote the rise of the quality of existing institutions
(especially nursing homes) that have to satisfy standard criteria
for care assistance and ({\em ii}) to implement a strategy to reform
the role of public institutions in this context. At this regard, a
great debate arises about the construction of indicators able to
measure nursing home performance and then to effectively rank the
facilities in a certain geographical region; see \cite{Phillips2007}
and \cite{harrington2003}. In the United States, one of the most
important projects of nursing home facility ranking is ``Medicare",
which is supported by the Department of Health and Human Services.
This project has developed an evaluation system  which is based on
indicators that are generally linked to the psycho-physics condition
of elderly people which are related, in particular, to mobility
diseases, behavioral disorders, and memory problems. Data are
collected by public institutions through questionnaires administered
at regularly repeated occasions.

One of the main ideas behind the facility ranking is that
unidimensional criteria to classify the health conditions of elderly
people are available; see \cite{Phillips2007} and
\cite{VincentMor04012003}. This assumption implies that the
difference between two subjects in responding to a set of items on
the health condition only depends on a single latent trait
summarizing this condition. This may obviously be a restrictive
assumption. For instance, this does not allow us to classify
subjects who show a degenerative health status in relation to a
specific pathology, but apparently have an overall good health
status. Violations of this assumption may lead to misleading
conclusions reached on the basis of a unidimensional ranking which
relies on a single score assigned to each facility.

On the other hand, \cite{kane1998} identified at list five different
groups of residents of nursing homes. These groups include subjects
recovering from an acute episode and are likely to return home,
residents who are cognitive impaired, residents who are cognitively
intact but suffer from physical challenges, residents who are in
vegetative state, and residents that are terminally ill. These
subjects have different needs and different levels of the quality of
life; see \cite{kane1981}. Consequently, the actions of the nursing
homes could be more forceful in relation to specific areas of
intervention, so that some facilities can be specialized in
improving the state of health in relation to some pathologies. Along
these lines, dimensionality of health condition becomes a central
issue for obtaining a consistent and generalizable ranking index for
the performance of the nursing homes.

In the present paper, we simultaneously address the issue of item
selection and of dimensionality on the basis of a formal statistical
procedure \citep{barolucci2007}, which exploits the latent class
(LC) model and a class of item response theory (IRT) models; see
\cite{laza:henr:68}, \cite{goodman1974, goodman1978}, and
\cite{Hambleton1996}. Through these methodologies, we study the
above issues on the basis of a dataset coming from the database
``Ulisse", which is collected within a survey carried out in Italy
on the basis of the RAI-MDS questionnaire \citep{Morris:1997}. The
questionnaire covers several aspects of the health status of elderly
people hosted in nursing homes. In particular, we consider 89 among
the around 300 available items. These items characterize: ({\em i})
cognitive conditions, ({\em ii}) auditory and view fields, ({\em
iii}) humor and behavioral disorders, ({\em iv}) activities of daily
living, ({\em v}) incontinence, ({\em vi}) nutritional field, ({\em
vii}) dental disorders, and ({\em viii}) skin conditions. By using a
so large number of items we can fully characterize the health status
of elderly people hosted in the nursing homes, without imposing any
\textit{a-priori} restriction on the relevance of its different
components. From the original set of 89 items we then extract a
subset of 35 items on the basis of their discriminant power, that is
the effectiveness in measuring these conditions. The adopted
methodology is based on the joint use of LC and IRT models and
represents a useful tool to reduce the size of the present and
similar questionnaires, with the obvious consequence of reducing the
survey costs.

On the basis of the applied methodology we show that the 35 selected
items indeed measure five different dimensions which may be referred
to as: ({\em i}) cognitive conditions, ({\em ii}) auditory and view
fields, ({\em iii}) activities of daily living and incontinence,
({\em iv}) humor and behavioral disorders and skin conditions, and
({\em v}) nutritional field and dental disorders. These dimensions
have a clear interpretation; this seems to confirm the robustness of
the proposed analysis.

The reminder of the paper is organized as follows. In Section 2 we
describe the dataset on health conditions of elderly people hosted
in certain of nursing homes across the Italian regions. In Section 3
we briefly review the statistical methodology based on LC and IRT
models. In Section 4 we describe in detail the empirical analysis
and in Section 5 we report the main conclusions of the study.
\section{The Ulisse database}
We consider a dataset collected within the ``Ulisse" project, which
is carried out by the Italian Ministry of Health jointly with the
Italian Society of Gerontology and Geriatrics. The project is based
on a longitudinal survey, covering 17 Italian regions, about the
assistance level provided to patients hosted in 37 randomly chosen
nursing homes. This survey is carried out since 2004 through the
repeated administration of a questionnaire (every 6 months) which is
filled up by the nursing assistant of each patient and concerns
several aspects of the everyday life. For our analysis we consider
only the first interview, which covers 1699 patients.

Table 1 reports the geographical distribution, on the Italian
territory, of the elderly people and the nursing homes included in
the study. We observe that most of the sample is in the north
regions: the percentage of patients in these regions is about
$85\%$. In particular, $40\%$ of the full sample is located in
Lombardia (17\%), Veneto  (11\%), and Emilia Romagna (12\%). In
Table 2 we report some descriptive statistics on gender and age of
patients of the nursing homes.

\begin{table}[h]
\caption{\em Regional distribution of the elderly people and nursing
homes included in the study.} \label{tabelone} \scriptsize{
\begin{tabularx}{\textwidth}{r X  r r r}
\toprule
    &             &   Number of &               & \multicolumn1c{Numbers of}\\
    &   Region    &   \multicolumn1c{subjects}  &   Percentage  & \multicolumn1c{nursing homes}\\
\hline
\\
North   &   Piemonte &   105 &   6.18    &   2    \\
    &   Lombardia&   292 &   17.19   &  6   \\
    &   Trentino A. A.  &   74  &   4.36    &   1   \\
    &   Veneto  &   194 &   11.42   &   5   \\
    &   Friuli V. G.    &   92  &   5.41    &   1  \\
    &   Liguria &   133  &   7.83    &   2  \\
    &   Emilia R.    &  214 &   12.60   &   4   \\
    Center& Tuscany &   83  &   4.89    &  3   \\
    &   Umbria  &   142 &   8.36    &   3  \\
    &   Marche  &   48  &   2.83    &  1   \\
South   &   Abruzzo &   78  &   4.59    &   2   \\
    &   Molise  &   46  &   2.71    &  2   \\
    &   Campania    &   52  &   3.06    &  1   \\
    &   Puglia  &   51  &   3.00    &  1 \\
    &   Calabria    &   42  &   2.47    &  1   \\
Ilands  &   Sicilia  &   31  &   1.82    &   1   \\
    &   Sardegna     &   22  &   1.29    &   1    \\
\hline
\\
    &   Total   &   1,699   &   100.00  &   37  \\
\bottomrule
\end{tabularx}}\vspace*{0.5cm}
\end{table}

\begin{table}[h]
\caption{\em Nursing homes population by age and gender.}
\scriptsize{
\begin{tabularx}{\textwidth}{X  r r r }
\toprule
Age &   Male  &   Female    &   Total   \\
\hline
\\
$>70    $&  4.72    &   3.24    &   7.96    \\
$70-75$ &   4.78    &   5.37    &   10.14   \\
$75-80$ &   6.07    &   11.85   &   17.92   \\
$80-85$ &   5.72    &   15.74   &   21.46   \\
$85-90$ &   3.36    &   15.68   &   19.04   \\
$>90$   &   4.19    &   19.28   &   23.47   \\
\hline
\\
Total   &   28.83   &   71.17   &   100.00  \\
\bottomrule
\end{tabularx}}\vspace*{0.5cm}
\end{table}

We observe that most of the sample is composed by women ($71\%$),
with only $29\%$ of men. Moreover, the age distribution differs in
relation to gender, with a higher proportion of females with age 85
and over and a relative younger male population. The presence of a
so high percentage of old women can obviously condition the analysis
about the care facility performance.

From the original questionnaire we single out 89 among the items
concerning: ({\em i}) cognitive conditions, ({\em ii}) auditory and
view fields, ({\em iii}) humor and behavioral disorders, ({\em iv})
activities of daily living, ({\em v}) incontinence, ({\em vi})
nutritional field, ({\em vii}) dental disorders, and ({\em viii})
skin conditions. The complete list of items  is reported in Appendix
1.
\section{The statistical methodology}
In the following, we briefly review the LC model \citep{goodman1974}
and the IRT model proposed by \cite{barolucci2007} for the study of
multidimensionality. The aim of the methodology based on these
models is: ({\em i}) to include in the analysis only the items
necessary to identify the number of latent traits;  ({\em ii}) to
identify the latent structure representing the health status of
elderly people; ({\em iii}) to investigate the different
discriminant power of the items. This methodology allows us to
choose a convenient partition of the selected items according to the
dimension they measure and to analyze its correspondence in terms of
facility care.
\subsection{The latent class model}\label{sec:LC}
The LC model \citep{laza:henr:68,goodman1974,goodman1978} assumes
that the observed sample is drawn from a population which is
partitioned into $k$ latent classes, with $\pi_c$ being the prior
probability (or weight) of class $c$ ($c=1,\ldots,k$). For each
subject $i$ ($i=1,\ldots,n$), we observe a vector $\b
y_{i}=(y_{i1},....,y_{iJ})$ of binary response variables
corresponding to $J$ items. Given that the subject is in class $c$
and with reference to item $j$, the conditional probability of
success is denoted by $\lambda_{j|c}$.

Under the assumption of {\em local independence}, the probability of
the response pattern $\b y_i$ is
\begin{eqnarray*}
p (\b y_{i})&=& \sum_cp(\b y_i|c)\pi_{c},\\\quad p(\b
y_i|c)&=&\prod_{j}\lambda_{j|c}^{y_{ij}}(1-\lambda_{j|c})^{1-y_{ij}}.
\end{eqnarray*}
The log-likelihood function of the LC model, which is used for
parameter estimation, is then
\begin{equation}
\ell(\b\theta)=\sum_i\log p(\b y_i)\label{eq:log-lik}
\end{equation}
where $\b\theta$ is a short-hand notation for all model parameters.
This function is maximized by the EM algorithm \citep{Dempster1977,
goodman1978}. This is an iterative algorithm which is based on two
steps to be repeated until convergence:
\begin{itemize}
\item {\bf E-step}: compute the conditional expected value of the
log-likelihood given the observed data and the current value of the
parameters;
\item {\bf M-step}: update the model parameters by maximizing the
expected log-likelihood obtained at the E-step.
\end{itemize}
We initialize the algorithm by both deterministic and random
starting values in order to prevent the problem of multimodality of
the likelihood. This is a typical problem of latent variable models.

Obviously, when the LC model is applied to analyze a dataset,
choosing the number of latent classes is necessary. At this aim, we
rely on the Bayesian Information Criterion (BIC) of
\cite{schwarz1978}, which is based on the index:
\[
BIC_k=-2\hat{\ell}_k+m_k\log(n),
\]
where, for a given number of classes $k$, $\hat{\ell}_k$ is the
maximum of the log-likelihood given in (\ref{eq:log-lik}) and $m_k$
is the corresponding number of parameters. The latter is taken as a
measure of complexity of the model on which the above penalization
term is based. According to this criterion, the number of classes
corresponding to the minimum of $BIC_k$ has to be selected. This
number is indicated by $\hat{k}$.

Through the EM algorithm we obtain, for each latent class $c$, the
maximum likelihood estimate of the weight, denoted by $\hat{\pi}_c$,
and of the conditional probabilities of success, denoted by
$\hat{\lambda}_{j|c}$, $j=1,\ldots,J$. On the basis of the latter
ones we can obtain a measure of the discriminant power of the items
measuring each dimension. In our study, each dimension corresponds
to a group of items measuring a different aspect of the health
status of elderly people. In particular, we measure the discriminant
power of item $j$ by the following ratio:
\begin{eqnarray}
D_j&=&\frac{M_j}
{\max_{h(j)}(M_h)},\label{eq:ratio}\\
M_j&=&\max_c(\hat{\lambda}_{j|c})-\min_c(\hat{\lambda}_{j|c}),\nonumber
\end{eqnarray}
where $\max_c(\hat{\lambda}_{j|c})$ is the maximum value (across the
latent classes) of the probability of success for item $j$ and
$\min_c(\hat{\lambda}_{j|c})$ is the minimum. Moreover,
$\max_{h(j)}$ at the denominator of (\ref{eq:ratio}) stands for the
maximum of the difference at the numerator with respect to all the
items measuring the same dimension of item $j$. In this way, the
above index is always between 0 and 1 and we exclude from the
analysis those items which have an index value lower than a given
threshold because they show a reduced discriminant power. This
approach for reducing the number of items is compared with the one
based on the IRT model that we present in the following section.
\subsection{The multidimensional two-parameter logistic
model}\label{sec:2pl}
This model is a constrained version of the LC model which directly
includes, for each item, a parameter measuring its discriminant
power. The model is based on the following multidimensional
two-parameter logistic (2PL) parametrization of the conditional
probabilities of success:
\begin{equation}
{\rm
logit}(\lambda_{j|c})=\gamma_{j}\left(\sum_d\delta_{jd}\theta_{cd}
-\beta_j\right), \qquad j=1,\ldots,J.\label{eq:2pl}
\end{equation}
In the above expression, $\delta_{jd}$ is a dummy variable equal to
1 if item $j$ measures dimension $d$ ($d=1,\ldots,s$) and to 0
otherwise. Moreover, $\theta_{cd}$ is a measure of the latent trait
(dimension $d$) for the subjects in latent class $c$ (typically
referred to as {\em ability}), $\beta_j$ is a measure of the overall
tendency to respond 0 to item $j$ (typically referred to as {\em
difficulty}), and $\gamma_j$ measures the discriminant power of this
item (typically referred to as {\em discriminant index}).

The log-likelihood function of the model may again be expressed as:
\[
\ell(\b\theta)=\sum_i\log\sum_{c}p(\b y_i|c)\pi_{c};
\]
maximization of $\ell(\b\theta)$ is again performed by the EM
algorithm of \cite{Dempster1977}, which has a the same structure
outlined in the previous section.

From the EM algorithm we also obtain, for each item $j$, the
estimate $\hat{\gamma}_j$ of the discriminant index. In order to
select a suitable set of items on the basis of these estimates, we
rely on the ratio
\begin{equation}
D_j^*=\frac{\hat{\gamma}_{j}}{\max_{h(j)}(\hat{\gamma}_{h})}.\label{eq:ratio2}
\end{equation}
This index has a structure similar to that in (\ref{eq:ratio}), with
the difference $M_j$ between the maximum and the minimum of the
estimated response probabilities substituted by the estimated
discriminant index. Then, items with a value of $D_j^*$ lower than a
certain threshold are dropped.

Another analysis that is allowed by the 2PL model presented above is
that of dimensionality. In particular, through this model we can
test the hypothesis, indicated in the following by $H_0$, that the
items actually measure a reduced number of dimensions. For instance,
we can test the hypothesis that the items measure $s-1$ instead of
$s$ dimensions. The $s-1$ dimensions are specified by collapsing two
dimensions into one and then grouping the corresponding items.

In order to test $H_0$, we use the likelihood ratio statistic
\[
LR=2\sum_{\bl y} n(\b y)\log \left[\frac{\hat{p}(\b y)}{\hat{p}_0(\b
y)}\right],
\]
where the sum is extended to all the possible response
configurations $\b y$, $n(\b y)$ stands for the observed frequency
of configuration $\b y$, $\hat{p}(\b y)$ is estimated probability of
this configuration under the model with $s$ dimensions, and
$\hat{p}_0(\b y)$ is the corresponding estimate under the reduced
model with $s-1$ dimensions. Under $H_0$, this statistic has a
$\chi^2$ asymptotic distribution with a number of degrees of freedom
equal to the $\hat{k}-2$, where $\hat{k}$ is the adopted number of
latent classes. Then, the hypothesis is rejected if the observed
value of $LR$ is larger than a suitable percentile of this
distribution.

On the basis of the above testing procedure, we can implement a
hierarchical algorithm for clustering items into a reduced number of
groups. Items in the same group are supposed to measure the same
dimension and then each group corresponds to a different dimension.
In particular, the clustering algorithm begins by fitting the model
in which items are grouped according the structure of the
questionnaire. Then, all the possible ways to collapse two groups
are considered and the one giving the smallest value of $LR$ with
respect to the previous step is selected. This procedure is repeated
until the unidimensional IRT model (in which all items are included
in the same group) is fitted. Finally, the selected number of groups
(and then of dimensions) is the smallest number for which the value
of $LR$, computed with respect to the previous classification, is
smaller than a suitable percentile of the asymptotic distribution.
For a detailed illustration of the algorithm see
\cite{barolucci2007}. We perform this classification once the items
with a reduced discriminating power are eliminated as indicated
above.
\section{Application to the Ulisse dataset}
Following the empirical strategy outlined in the previous section,
we analyze the Ulisse dataset described in Section 2. We first
present the results from the LC model and then those from the IRT
model.
\subsection{Latent class analysis}
As starting point we choose the number of latent classes for the LC
model applied to the 89 selected items. At this aim, Table \ref{bic}
reports the maximum log-likelihood and the corresponding value of
the BIC index for a number of latent classes from 1 to 7.
\begin{table}[ht]\centering
\caption{\em Selection of the number of latent classes for the LC
model. For each number of classes from 1 to 7, $\hat{\ell}_k$ is the
maximum log-likelihood of the model, $m_k$ is the number of
parameters, and $BIC_k$ is the corresponding value of the BIC index.
In boldface are reported the quantities corresponding to the
selected model.} {\small
\begin{tabular}{r r r r }
        \toprule
    $k$& \multicolumn1c{$\hat{\ell}_k$} & \multicolumn1c{$m_k$} & \multicolumn1c{$BIC_k$}\\
    \hline \\
1   &   -37175.939  &   89  &   75013.842\\
2   &   -32773.208  &   179 &   66877.781\\
3   &   -31444.523  &   269 &   64889.813\\
4   &   -30432.381  &   359 &   63534.930\\
5   &   -29875.393  &   449 &   63090.356\\
{\bf 6}   &   {\bf -29423.379}  &   {\bf 539} &   {\bf 62855.730}   \\
7   &   -29159.367  &   629 &   62997.107\\
\bottomrule
\end{tabular}}
\label{bic}\vspace*{0.5cm}
\end{table}

\begin{figure}[ht]
   \caption{\em Plot of the estimated conditional probabilities
   of success $\hat{\lambda}_{j|c}$ under the selected LC model.
   Each panel corresponds to a different latent class $c$, ordered
   according to the probability of success for the first item.}
   \centering
   \includegraphics[height=12.5cm]{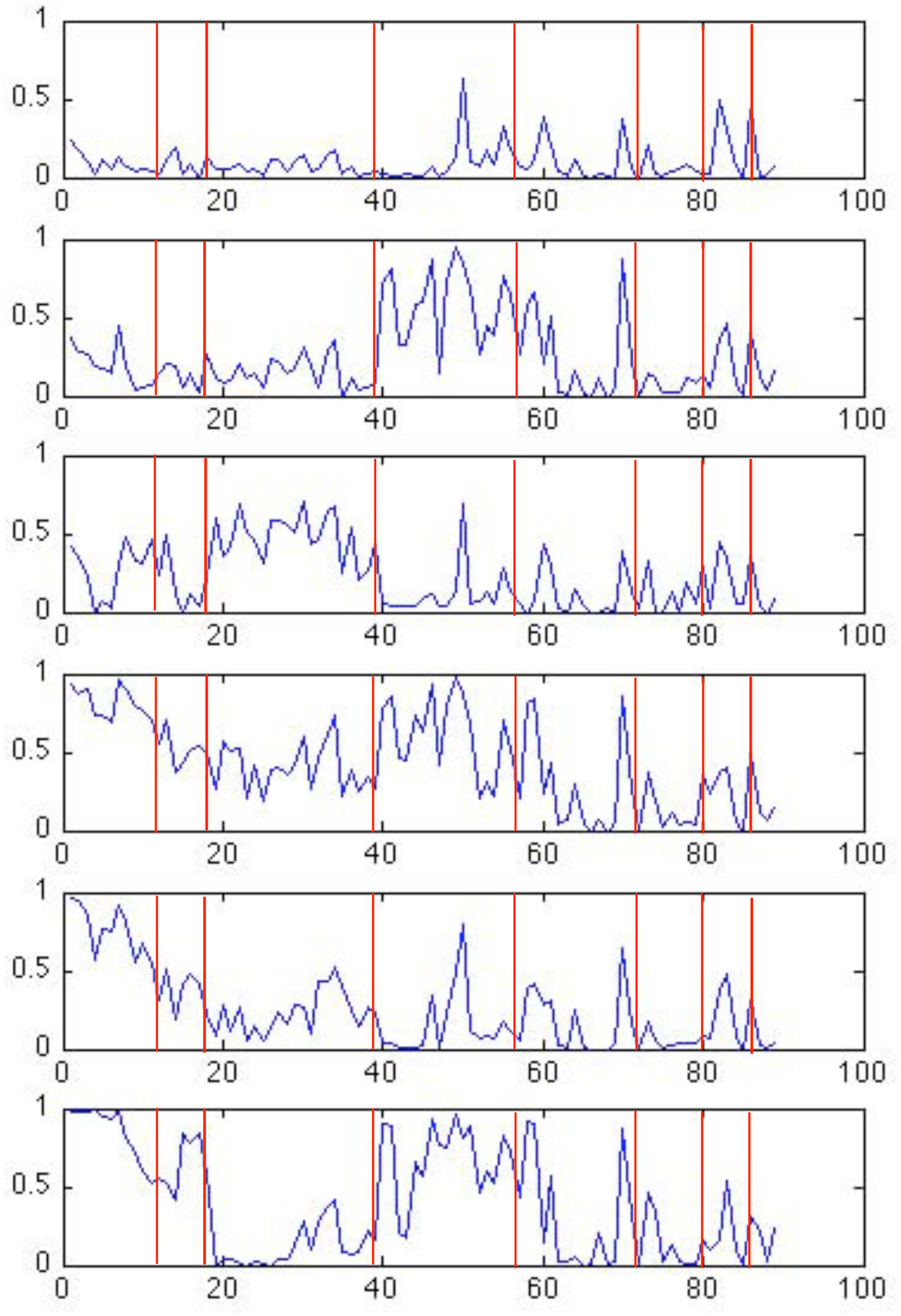}
   \label{fig1}
\end{figure}

On the basis of these results we select $\hat{k}=6$ latent classes.
These classes correspond to different degrees of impairment of the
elderly people health status. In order to interpret these classes,
Figure \ref{fig1} reports the graph of the conditional probabilities
of success estimated under the selected LC model for each class. To
have a clearer interpretation of these results, we order the six
classes according to the probability of success for the first item.

\newpage

The estimated conditional probabilities obtained from the LC model
are then used to assess the discriminant power of the 89 selected
items. At this aim, in Table \ref{tab5} we report for each item $j$
the value of the index $D_j$, computed according to
(\ref{eq:ratio}), together with the weighted mean and standard
deviation of the estimated success probabilities, computed with
weights corresponding to the estimated class probabilities.\newpage

\begin{table}[h]\centering
\caption{\em Weighted mean and standard deviation of the estimated
success probabilities $\hat{\lambda}_{j|c}$ for each item $j$,
together with the indices $M_j$ and $D_j$ used to measure the
discriminant power. In boldface are the quantities referred to the
item that for each dimension $d$ has the highest discriminant
power.} \tiny{\
\begin{tabular}{rlcrrrrcrlcrrrr}
        \toprule
\multicolumn1c{$j$}  &   item    &   $d$   &   \multicolumn1c{mean}    &   \multicolumn1c{std} &   \multicolumn1c{$M_j$}  &   \multicolumn1c{$D_j$}     &&
\multicolumn1c{$j$}  &   item    &   $d$   &   \multicolumn1c{mean}    &   \multicolumn1c{std} &   \multicolumn1c{$M_j$}  &   \multicolumn1c{$D_j$}\\   \hline \\
1  &   CC1 &   1   &   0.629   &   0.107   &   0.753   &   0.760       &&  \b{40}& \b{ADL1}    &   \b{4}   &   \b{0.459}   &   \b{0.243}   &   \b{0.876}   &   \b{1.000}   \\
2  &   CC2 &   1   &   0.577   &   0.126   &   0.807   &   0.814       &&  41  &   ADL2    &   4   &   0.481   &   0.264   &   0.869   &   0.992   \\
3  &   CC3 &   1   &   0.538   &   0.177   &   0.875   &   0.883       &&  42  &   ADL3    &   4   &   0.202   &   0.289   &   0.461   &   0.526   \\
\b{4}   &   \b{CC4} &   \b{1}   &   \b{0.404}   &   \b{0.277}   &   \b{0.991}   &   \b{1.000}       &&  43  &   ADL4    &   4   &   0.198   &   0.290   &   0.440   &   0.502   \\
5  &   CC5 &   1   &   0.442   &   0.226   &   0.877   &   0.885       &&  44  &   ADL5    &   4   &   0.382   &   0.241   &   0.721   &   0.823   \\
6  &   CC6 &   1   &   0.398   &   0.264   &   0.921   &   0.929       &&  45  &   ADL6    &   4   &   0.355   &   0.214   &   0.618   &   0.705   \\
7  &   CC7 &   1   &   0.609   &   0.142   &   0.865   &   0.873       &&  46  &   ADL7    &   4   &   0.563   &   0.215   &   0.870   &   0.993   \\
8  &   CC8 &   1   &   0.542   &   0.119   &   0.827   &   0.835       &&  47  &   ADL8    &   4   &   0.246   &   0.291   &   0.753   &   0.860   \\
9  &   CC9 &   1   &   0.426   &   0.159   &   0.761   &   0.768       &&  48  &   ADL9    &   4   &   0.449   &   0.224   &   0.754   &   0.861   \\
10  &   CC10    &   1   &   0.410   &   0.154   &   0.711   &   0.717       &&  49  &   ADL10   &   4   &   0.618   &   0.207   &   0.857   &   0.978   \\
11  &   CC11    &   1   &   0.401   &   0.126   &   0.668   &   0.674       &&  50  &   ADL11   &   4   &   0.775   &   0.030   &   0.238   &   0.272   \\
12  &   CC12    &   1   &   0.316   &   0.173   &   0.533   &   0.538       &&  51  &   ADL12   &   4   &   0.442   &   0.214   &   0.830   &   0.947   \\
13  &   CC13    &   1   &   0.439   &   0.086   &   0.593   &   0.598       &&  52  &   ADL13   &   4   &   0.187   &   0.279   &   0.395   &   0.451   \\
\cline{1-7}
14  &   CAS1    &   2   &   0.254   &   0.205   &   0.295   &   0.348       &&  53  &   ADL14   &   4   &   0.299   &   0.189   &   0.524   &   0.598   \\
\b{15}  &   \b{CAS2}    &   \b{2}   &   \b{0.277}   &   \b{0.296}   &   \b{0.847}   &   \b{1.000}       &&  54  &   ADL15   &   4   &   0.224   &   0.262   &   0.480   &   0.548   \\
16  &   CAS3    &   2   &   0.341   &   0.202   &   0.704   &   0.831       &&  55  &   ADL16   &   4   &   0.538   &   0.092   &   0.652   &   0.744   \\
17  &   CAS4    &   2   &   0.302   &   0.282   &   0.843   &   0.995       &&  56  &   ADL17   &   4   &   0.401   &   0.159   &   0.611   &   0.697   \\
18  &   CAS5    &   2   &   0.331   &   0.152   &   0.465   &   0.549       &&  57  &   ADL18   &   4   &   0.183   &   0.285   &   0.378   &   0.432   \\
\cline{1-7}\cline{9-15}
19  &   HBD1    &   3   &   0.220   &   0.212   &   0.598   &   0.882       &&  \b{58}  &   \b{I1}  &   \b{5}   &   \b{0.462}   &   \b{0.245}   &   \b{0.917}   &   \b{1.000}   \\
20  &   HBD2    &   3   &   0.262   &   0.195   &   0.532   &   0.785       &&  59  &   I2  &   5   &   0.512   &   0.194   &   0.821   &   0.895   \\
21  &   HBD3    &   3   &   0.251   &   0.194   &   0.482   &   0.711       &&  60  &   I3  &   5   &   0.288   &   0.135   &   0.294   &   0.321   \\
\b{22}  &   \b{HBD4}    &   \b{3}   &   \b{0.335}   &   \b{0.150}   &   \b{0.678}   &   \b{1.000}       &&  61  &   I4  &   5   &   0.392   &   0.093   &   0.353   &   0.385   \\
23  &   HBD5    &   3   &   0.182   &   0.235   &   0.506   &   0.746       &&  62  &   I5  &   5   &   0.026   &   0.440   &   0.020   &   0.022   \\
24  &   HBD6    &   3   &   0.243   &   0.187   &   0.441   &   0.650       &&  63  &   I6  &   5   &   0.022   &   0.450   &   0.061   &   0.067   \\
25  &   HBD7    &   3   &   0.126   &   0.287   &   0.308   &   0.454       &&  64  &   I7  &   5   &   0.174   &   0.261   &   0.251   &   0.274   \\
26  &   HBD8    &   3   &   0.283   &   0.161   &   0.588   &   0.867       &&  65  &   I8  &   5   &   0.019   &   0.445   &   0.038   &   0.041   \\
27  &   HBD9    &   3   &   0.295   &   0.150   &   0.555   &   0.819       &&  66  &   I9  &   5   &   0.002   &   0.474   &   0.011   &   0.012   \\
28  &   HBD10   &   3   &   0.249   &   0.185   &   0.524   &   0.773       &&  67  &   I10 &   5   &   0.068   &   0.401   &   0.200   &   0.218   \\
29  &   HBD11   &   3   &   0.292   &   0.143   &   0.409   &   0.603       &&  68  &   I11 &   5   &   0.006   &   0.465   &   0.016   &   0.017   \\
30  &   HBD12   &   3   &   0.424   &   0.088   &   0.568   &   0.838       &&  69  &   I12 &   5   &   0.026   &   0.444   &   0.054   &   0.059   \\
31  &   HBD13   &   3   &   0.209   &   0.203   &   0.393   &   0.580       &&  70  &   I13 &   5   &   0.665   &   0.068   &   0.499   &   0.544   \\
32  &   HBD14   &   3   &   0.297   &   0.150   &   0.423   &   0.624       &&  71  &   I14 &   5   &   0.197   &   0.245   &   0.281   &   0.306   \\
33  &   HBD15   &   3   &   0.443   &   0.070   &   0.502   &   0.740       &&  72  &   I15 &   5   &   0.007   &   0.464   &   0.016   &   0.017   \\
\cline{9-15}
34  &   HBD16   &   3   &   0.503   &   0.057   &   0.553   &   0.816       &&  73  &   N1  &   6   &   0.296   &   0.140   &   0.304   &   0.910   \\
35  &   HBD17   &   3   &   0.150   &   0.273   &   0.396   &   0.584       &&  \b{74}  &   \b{N2}  &   \b{6}   &   \b{0.121}   &   \b{0.352}   &   \b{0.334}   &   \b{1.000}   \\
36  &   HBD18   &   3   &   0.264   &   0.170   &   0.481   &   0.709       &&  75  &   N3  &   6   &   0.014   &   0.459   &   0.030   &   0.090   \\
37  &   HBD19   &   3   &   0.132   &   0.293   &   0.245   &   0.361       &&  76  &   N4  &   6   &   0.087   &   0.347   &   0.105   &   0.314   \\
38  &   HBD20   &   3   &   0.204   &   0.223   &   0.322   &   0.475       &&  77  &   N5  &   6   &   0.026   &   0.445   &   0.051   &   0.153   \\
39  &   HBD21   &   3   &   0.217   &   0.197   &   0.396   &
0.584       &&  78  &   N6  &   6   &   0.088   &   0.335   &
0.182   &   0.545   \\\cline{1-7}
    &       &       &       &       &       &           &&  79  &   N7  &   6   &   0.043   &   0.412   &   0.072   &   0.216   \\
    &       &       &       &       &       &           &&  80  &   N8  &   6   &   0.203   &   0.220   &   0.329   &   0.985   \\
\cline{9-15}
    &       &       &       &       &       &           &&  81  &   D1  &   7   &   0.094   &   0.370   &   0.225   &   0.674   \\
    &       &       &       &       &       &           &&  \b{82}  &   \b{D2}  &   \b{7}   &   \b{0.377}   &   \b{0.081}   &   \b{0.334}   &   \b{1.000}   \\
    &       &       &       &       &       &           &&  83  &   D3  &   7   &   0.409   &   0.072   &   0.238   &   0.713   \\
    &       &       &       &       &       &           &&  84  &   D4  &   7   &   0.068   &   0.383   &   0.047   &   0.141   \\
    &       &       &       &       &       &           &&  85  &   D5  &   7   &   0.013   &   0.449   &   0.049   &   0.147   \\
    &       &       &       &       &       &           &&  86  &   D6  &   7   &   0.398   &   0.073   &   0.179   &   0.536   \\
\cline{9-15}
    &       &       &       &       &       &           &&  \b{87}  &   \b{SK1} &   \b{8}   &   \b{0.088}   &   \b{0.370}   &   \b{0.220}   &   \b{1.000}   \\
    &       &       &       &       &       &           &&  88  &   SK2 &   8   &   0.029   &   0.441   &   0.062   &   0.282   \\
    &       &       &       &       &       &           &&  89  &   SK3 &   8   &   0.126   &   0.317   &   0.203   &   0.923   \\
\bottomrule
\end{tabular}}
\label{tab5}\vspace*{0.5cm}
\end{table}

As described in Section \ref{sec:LC}, the index $D_j$ may be used to
select a subset of items which provide a similar amount of
information than the full set of items. This is obtained by
comparing the values of this index with a suitable threshold between
0 and 1. In particular, for different threshold levels we report in
Table \ref{tab2} the number of selected items for each dimension.
For instance, with a threshold of $0.5$ we retain in the analysis 63
items.

\begin{table}[h]\centering
\caption{\em Results from the item selection procedure based on the
indices $D_j$ in terms of number of items retained for each
dimension.} {\small
\begin{tabular}{crrrrrrrrc}
\toprule & \multicolumn8c{dimension}\\\cline{2-9} \\threshold &
\multicolumn1c{1} & \multicolumn1c{2} & \multicolumn1c{3} &
\multicolumn1c{4} & \multicolumn1c{5} &
\multicolumn1c{6} & \multicolumn1c{7} & \multicolumn1c{8} &  overall\\
\hline \\
0.0&13&5&21&17&16&8&6&3&89\\
0.1&13&5&21&18&8&7&6&3&81\\
0.2&13&5&21&18&8&6&4&3&78\\
0.3&13&5&21&17&6&5&4&2&73\\
0.4&13&4&20&17&3&4&4&2&67\\
0.5&13&4&18&15&3&4&4&2&63\\
0.6&11&3&15&11&2&3&3&2&50\\
0.7&10&3&12&10&2&3&2&2&44\\
0.8&7&3&6&8&2&3&1&2&32\\
0.9&2&2&1&5&1&3&1&2&17\\
1.0&1&1&1&1&1&1&1&1&$\:\:$1\\
\bottomrule
\end{tabular}}
\label{tab2}\vspace*{0.5cm}
\end{table}
\newpage
\subsection{Item response theory analysis }
To complete the item selection analysis we exploit the alternative
approach based on the 2PL model illustrated in Section
\ref{sec:2pl}. The same model is then used to evaluate the
dimensionality of health condition of elderly people. The results of
the analysis are finally compared with those presented in the
previous section and based on the LC model. Since the adopted IRT
model can be seen as a constrained version of the LC model, there
are no compatibility problems in comparing the results of the two
analyses.

On the basis of the same number of classes selected above,
$\hat{k}=6$, we obtain the estimates of the parameters in
(\ref{eq:2pl}). In particular, for each item $j$ we report in Table
\ref{table3} the estimated difficulty level $\hat{\beta}_j$ and
discriminant index $\hat{\gamma}_j$, together with the index $D_j^*$
defined in (\ref{eq:ratio2}). On the basis of the values of this
index, we can select a suitable number of items. The results of this
selection process are reported in Table \ref{tableirt2} for
different threshold levels between  0 to 1.

\begin{table}[h]\centering
\caption{\em Parameter estimates under the 2PL model. For each item
$j$, $\hat{\beta}_j$ is the estimated difficulty, $\hat{\gamma}_j$
is the estimated discriminant power, and $D_j^*$ is the relative
discriminant power. In boldface are the quantities referred to the
item that for each dimension $d$ has the highest discriminant
power.} \tiny{
\begin{tabular}{crlrrrccrlrrr}
\toprule $d$   &   \multicolumn1c{$j$}  & \multicolumn1c{item} &
\multicolumn1c{$\hat{\gamma}_j$} & \multicolumn1c{$\hat{\beta}_j$} &
\multicolumn1c{$D_j^*$}  && $d$   &   \multicolumn1c{$j$}  &
\multicolumn1c{item} & \multicolumn1c{$\hat{\gamma}_j$}    &
\multicolumn1c{$\hat{\beta}_j$} &   \multicolumn1c{$D_j^*$} \\
\hline \\
1   &   1   &   CC1 &   1.000   &   0.000   &   0.706   &&  4   &   40  &   ADL1    &   1.000   &   0.000   &   0.819   \\
1   &   2   &   CC2 &   1.161   &   0.492   &   0.820   &&  4   &   41  &   ADL2    &   1.107   &   -0.142  &   0.907   \\
\b{1}   &   \b{3}   &   \b{CC3} &   \b{1.416}   &   \b{0.875}   &   \b{1.000}   &&  4   &   42  &   ADL3    &   0.681   &   2.843   &   0.558   \\
1   &   4   &   CC4 &   1.178   &   1.742   &   0.832   &&  4   &   43  &   ADL4    &   0.585   &   3.123   &   0.479   \\
1   &   5   &   CC5 &   0.989   &   1.347   &   0.699   &&  4   &   44  &   ADL5    &   1.018   &   0.872   &   0.834   \\
1   &   6   &   CC6 &   1.098   &   1.666   &   0.775   &&  4   &   45  &   ADL6    &   0.817   &   1.082   &   0.669   \\
1   &   7   &   CC7 &   1.280   &   0.283   &   0.904   &&  4   &   46  &   ADL7    &   0.975   &   -1.303  &   0.799   \\
1   &   8   &   CC8 &   0.938   &   1.134   &   0.663   &&  4   &   \b{47}  &   \b{ADL8}    &   \b{1.221}   &   \b{1.758}   &   \b{1.000}   \\
1   &   9   &   CC9 &   0.928   &   2.093   &   0.655   &&  4   &   48  &   ADL9    &   0.732   &   -0.051  &   0.600   \\
1   &   10  &   CC10    &   0.805   &   2.154   &   0.568   &&  4   &   49  &   ADL10   &   1.155   &   -2.084  &   0.946   \\
1   &   11  &   CC11    &   0.750   &   2.501   &   0.530   &&  4   &   50  &   ADL11   &   0.212   &   -6.810  &   0.174   \\
1   &   12  &   CC12    &   0.657   &   3.113   &   0.464   &&  4   &   51  &   ADL12   &   0.641   &   -0.132  &   0.525   \\
1   &   13  &   CC13    &   0.463   &   2.355   &   0.327   &&  4   &   52  &   ADL13   &   0.386   &   3.596   &   0.316   \\
\cline{1-6}
2   &   14  &   CAS1    &   1.000   &   0.000   &   0.177   &&  4   &   53  &   ADL14   &   0.337   &   2.049   &   0.276   \\
2   &   15  &   CAS2    &   4.954   &   -0.806  &   0.879   &&  4   &   54  &   ADL15   &   0.429   &   2.626   &   0.352   \\
2   &   16  &   CAS3    &   3.673   &   -0.883  &   0.652   &&  4   &   55  &   ADL16   &   0.415   &   -1.089  &   0.340   \\
\b{2}   &   \b{17}  &   \b{CAS4}    &   \b{5.636}   &   \b{-0.829}  &   \b{1.000}   &&  4   &   56  &   ADL17   &   0.461   &   0.418   &   0.378   \\
2   &   18  &   CAS5    &   1.702   &   -0.602  &   0.302   &&  4   &   57  &   ADL18   &   0.392   &   3.674   &   0.321   \\
\cline{1-6} \cline{8-13}
3   &   19  &   HBD1    &   1.000   &   0.000   &   0.070   &&  5   &   58  &   I1  &   1.000   &   0.000   &   0.050   \\
3   &   20  &   HBD2    &   5.773   &   -1.646  &   0.403   &&  5   &   59  &   I2  &   1.013   &   -0.349  &   0.051   \\
3   &   21  &   HBD3    &   3.630   &   -1.453  &   0.253   &&  5   &   60  &   I3  &   0.086   &   10.000  &   0.004   \\
3   &   22  &   HBD4    &   2.722   &   -1.459  &   0.190   &&  5   &   61  &   I4  &   0.327   &   1.181   &   0.016   \\
3   &   23  &   HBD5    &   0.508   &   2.305   &   0.035   &&  5   &   62  &   I5  &   0.355   &   10.000  &   0.018   \\
3   &   24  &   HBD6    &   2.959   &   -1.352  &   0.206   &&  5   &   63  &   I6  &   0.654   &   6.530   &   0.033   \\
3   &   25  &   HBD7    &   3.225   &   -1.110  &   0.225   &&  5   &   64  &   I7  &   0.160   &   10.000  &   0.008   \\
3   &   26  &   HBD8    &   1.111   &   -0.626  &   0.078   &&  5   &   65  &   I8  &   0.434   &   10.000  &   0.022   \\
3   &   27  &   HBD9    &   1.867   &   -1.208  &   0.130   &&  \b{5}   &   \b{66}  &   \b{I9}  &   \b{20.000}  &   \b{1.914}   &   \b{1.000}   \\
3   &   28  &   HBD10   &   3.443   &   -1.425  &   0.240   &&  5   &   67  &   I10 &   0.586   &   4.627   &   0.029   \\
3   &   29  &   HBD11   &   3.885   &   -1.586  &   0.271   &&  5   &   68  &   I11 &   0.541   &   10.000  &   0.027   \\
3   &   30  &   HBD12   &   3.285   &   -1.693  &   0.229   &&  5   &   69  &   I12 &   0.421   &   8.777   &   0.021   \\
3   &   31  &   HBD13   &   2.297   &   -1.144  &   0.160   &&  5   &   70  &   I13 &   0.605   &   -1.818  &   0.030   \\
3   &   32  &   HBD14   &   12.159  &   -1.787  &   0.848   &&  5   &   71  &   I14 &   0.436   &   3.367   &   0.022   \\
3   &   33  &   HBD15   &   5.465   &   -1.813  &   0.381   &&  5   &   72  &   I15 &   0.545   &   10.000  &   0.027   \\

 \cline{8-13}
3   &   34  &   HBD16   &   6.356   &   -1.869  &   0.443   &&  6   &   73  &   N1  &   1.000   &   0.000   &   0.359   \\
\b{3}   &   \b{35}  &   \b{HBD17}   &   \b{14.337}  &   \b{-1.729}  &   \b{1.000}   &&  \b{6}   &   \b{74}  &   \b{N2}  &   \b{2.784}   &   \b{-0.142}  &   \b{1.000}   \\
3   &   36  &   HBD18   &   3.580   &   -1.494  &   0.250   &&  6   &   75  &   N3  &   0.975   &   3.359   &   0.350   \\
3   &   37  &   HBD19   &   8.130   &   -1.606  &   0.567   &&  6   &   76  &   N4  &   1.267   &   1.132   &   0.455   \\
3   &   38  &   HBD20   &   9.393   &   -1.713  &   0.655   &&  6   &   77  &   N5  &   0.310   &   10.000  &   0.111   \\
3   &   39  &   HBD21   &   4.940   &   -1.568  &   0.345   &&  6   &   78  &   N6  &   0.233   &   10.000  &   0.084   \\
\cline{1-7}
        &   &       &       &       &       &&  6   &   79  &   N7  &   0.289   &   10.000  &   0.104   \\
        &   &       &       &       &       &&  6   &   80  &   N8  &   1.350   &   0.365   &   0.485   \\
         \cline{8-13}
        &   &       &       &       &       &&  \b{7}   &   \b{81}  &   \b{D1}  &   \b{1.000}   &   \b{0.000}   &   \b{1.000}   \\
        &   &       &       &       &       &&  7   &   82  &   D2  &   0.050   &   8.553   &   0.050   \\
        &   &       &       &       &       &&  7   &   83  &   D3  &   0.583   &   -2.037  &   0.583   \\
        &   &       &       &       &       &&  7   &   84  &   D4  &   0.201   &   10.000  &   0.201   \\
        &   &       &       &       &       &&  7   &   85  &   D5  &   0.710   &   4.550   &   0.710   \\
        &   &       &       &       &       &&  7   &   86  &   D6  &   0.050   &   6.764   &   0.050   \\
         \cline{8-13}
        &   &       &       &       &       &&  \b{8}   &   \b{87}  &   \b{SK1} &   \b{1.000}   &   \b{0.000}   &   \b{1.000}   \\
        &   &       &       &       &       &&  8   &   88  &   SK2 &   0.328   &   8.156   &   0.328   \\
        &   &       &       &       &       &&  8   &   89  &   SK3 &   0.519   &   1.185   &   0.519   \\
        \bottomrule
\end{tabular}}
\label{table3}\vspace*{0.5cm}
\end{table}

\begin{table}[h!]\centering
\caption{\em Results from the item selection mechanism based on the
indices $D_j^*$ in terms of number of items retained for each
dimension.} {\small
\begin{tabular}{crrrrrrrrc}
\toprule & \multicolumn8c{dimension}\\\cline{2-9} \\threshold &
\multicolumn1c{1} & \multicolumn1c{2} & \multicolumn1c{3} &
\multicolumn1c{4} & \multicolumn1c{5} &
\multicolumn1c{6} & \multicolumn1c{7} & \multicolumn1c{8} &  overall\\
\hline \\
0.0 &   13  &   5   &   21  &   18  &   15  &   8   &   6   &   3   &   89  \\
0.1 &   13  &   5   &   18  &   18  &   1   &   7   &   4   &   3   &   69  \\
0.2 &   13  &   4   &   15  &   17  &   1   &   5   &   4   &   3   &   62  \\
0.3 &   13  &   4   &   8   &   16  &   1   &   5   &   3   &   3   &   53  \\
0.4 &   12  &   3   &   6   &   11  &   1   &   3   &   3   &   2   &   41  \\
0.5 &   11  &   3   &   4   &   10  &   1   &   1   &   3   &   2   &   35  \\
0.6 &   9   &   3   &   3   &   8   &   1   &   1   &   2   &   1   &   28  \\
0.7 &   6   &   2   &   2   &   6   &   1   &   1   &   2   &   1   &   21  \\
0.8 &   4   &   2   &   2   &   5   &   1   &   1   &   1   &   1   &   17  \\
0.9 &   2   &   1   &   1   &   3   &   1   &   1   &   1   &   1   &   11  \\
1.0 &   1   &   1   &   1   &   1   &   1   &   1   &   1   &   1   &   $\:\:$8   \\
\bottomrule
\end{tabular}}
\label{tableirt2}
\end{table}
\newpage

Compared to the LC model (see Table \ref{tab2}), the IRT model
chooses a reduced number of items to keep into the analysis,
appearing less conservative in terms of the item selection process.
In fact, by using a critical value of $0.5$ we select 35 items,
instead of 63 chosen by the LC based procedure. The choice of that
critical value has been addressed to keep in the analysis a relevant
number of items, without loosing too much information in relation to
the analyzed phenomenon.\newpage

\begin{figure}[h]\centering
\caption{\em Dendrogram based on the 2PL multidimensional model. The
eight initial dimensions characterize: (i) cognitive conditions,
(ii) auditory and view fields, (iii) humor and behavioral disorders,
(iv) activities of daily living, (v) continence, (vi) nutritional
field, (vii) dental disorders, and (viii) skin conditions.}
\includegraphics[height=10.25cm]{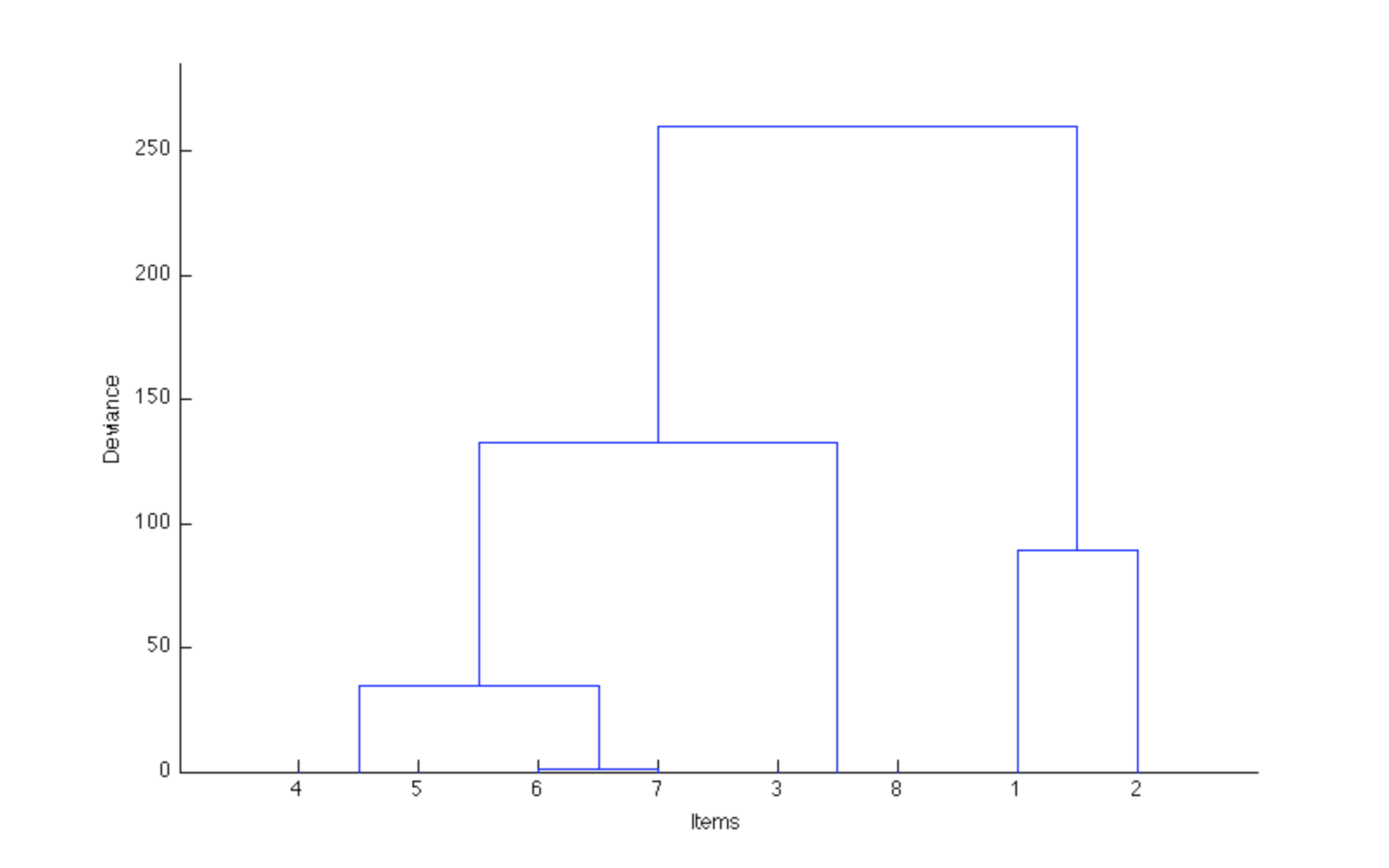}
\label{figure1}
\end{figure}

Once we have selected the number of items, we perform the
hierarchical cluster analysis on the dimensionality, starting from
the eight dimensions defined by the structure of the questionnaire.
The approach that we use at this aim is described at the end of
Section \ref{sec:2pl} and the results obtained from its application
are reported in Table \ref{tab111} and represented by the dendrogram
in Figure \ref{figure1}. In particular, the table shows the list of
the dimensions formed by the collection of the items corresponding
to the eight initial dimensions, together with the statistic $LR$
computed with respect to the model chosen at the previous step and
the corresponding $p$-value.

\begin{table}[h]\centering
\caption{\em Output of the hierarchical cluster algorithm based on
the 2PL multidimensional model.} {\small
\begin{tabular}{rrlrr}
\toprule $h$ & $s$ & clusters & \multicolumn1c{$LR$} & $p$-value      \\\hline\\
1   &   7   &   $\{1\},\{2\},\{3\},\{6\},\{7\},\{8\},\{4,5\}$   &  0.379 &  0.984\\
2   &   6   &   $\{1\},\{2\},\{6\},\{7\},\{4,5\},\{3,8\}$   & 3.871&   0.424\\
3   &   5   &   $\{1\},\{2\},\{4,5\},\{3,8\},\{6,7\}$   &5.235  &0.264\\
4   &   4   &   $\{1\},\{2\},\{3,8\},\{4,5,6,7\}$ &29.045&0.000\\
5   &   3   &   $\{3,8\},\{4,5,7,8\},\{1,2\}$   &   60.142  &   0.000    \\
6   &   2   &   $\{1,2\},\{3,4,5,6,7,8\}$ &   42.107  &   0.000    \\
7   &   1   &   $\{1,2,3,4,5,6,7,8\}$&   187.440 &   0.000    \\
\bottomrule
\end{tabular}}
\label{tab111}\vspace*{0.5cm}
\end{table}\newpage

These results show evidence of five dimensions which have the
following structure: $\{1\},\{2\},\{4,5\},\{3,8\},\{6,7\}$. These
dimensions correspond to: ({\em i}) cognitive conditions, ({\em ii})
auditory and view fields, ({\em iii}) activities of daily living and
incontinence, ({\em iv}) humor and behavioral disorders and skin
conditions, and ({\em v}) nutritional field and dental disorders.
These dimensions have a clear interpretation and seems to confirm
the robustness of the proposed analysis.

Finally, in Table \ref{tab:theta} we report the estimated abilities
$\hat{\theta}_{cd}$ for each latent class, together with estimated
class weights $\hat{\pi}_c$. With reference to every dimension $d$,
each parameter $\theta_{cd}$ corresponds to the latent trait level
for the subjects in class $c$. In the present study, high values of
the parameter correspond to high probability to suffer from a
certain pathology.

\begin{table}[h]\centering
\caption{\em Estimated ability parameters $\hat{\theta}_{cd}$ for
each latent class $c$ and dimension $d$, together with the estimated
weights $\hat{\pi}_c$.} {\small
\begin{tabular}{crrrrrr}
\toprule latent      & \multicolumn5c{dimension}\\\cline{2-6} class
& \multicolumn1c{1} & \multicolumn1c{2} &
\multicolumn1c{3} & \multicolumn1c{4} & \multicolumn1c{5}& weight \\\hline\\
1 & -2.516   &   -2.690  &   -5.179  &   -5.137  &   -4.815  &0.213\\
2 & 1.142    &   -1.171  &   -3.402  &   -2.224  &   -3.927& 0.153\\
3 & -1.253   &   -2.711  &   -2.227  &   -3.203  &   1.050&0.131 \\
4 & 3.996    &   0.702   &   -1.960  &   0.525   &   -2.025&0.102    \\
5 & 2.068    &   -1.667  &   -1.946  &   -1.973  &   1.333   &0.160\\
6 & 4.386    &   -0.451  &   -0.727  &   2.117   &   1.787&0.238
\\\hline
\end{tabular}}\label{tab:theta}\vspace*{0.5cm}
\end{table}

For each pair of dimensions $(d_1,d_2)$, it may be interesting to
compute the correlation $\rho_{d_1,d_2}$ between the estimated
ability levels. Taking into account the class weights, we compute
these correlation indices as
\[
\rho_{d_1d_2} = \frac{\sum_c
(\hat{\theta}_{cd_1}-\hat{\bar{\theta}}_{d_1})(\hat{\theta}_{cd_2}-\hat{\bar{\theta}}_{d_2})\hat{\pi}_c}
{\sqrt{\sum_c
(\hat{\theta}_{cd_1}-\hat{\bar{\theta}}_{d_1})\hat{\pi}_c}\sqrt{\sum_c
(\hat{\theta}_{cd_2}-\hat{\bar{\theta}}_{d_2})\hat{\pi}_c}},
\]
where $\hat{\bar{\theta}}_d=\sum_c \hat{\theta}_{cd}\hat{\pi}_c$ is
the average ability level for dimension $d$. The results are
reported in Table \ref{tabdsd}.

\begin{table}[h]\centering\vspace*{0.5cm}
\caption{\em Correlation between the estimated abilities for each
pair of dimensions ($d_1,d_2$).} {\small
\begin{tabular}{cccccc}\hline
& \multicolumn{5}{c}{1st dimension}\\\cline{2-6}
2nd dimension    &   1  &   2  &   3  &   4  & 5     \\\hline\\
1  &   1.000   \\
2   &   0.912   &   1.000   \\
3   &   0.866   &   0.648   &   1.000  \\
4   &   0.964   &   0.863   &   0.898   &   1.000   \\
5   &   0.611   &   0.287   &   0.905   &   0.661   &   1.000   \\
\bottomrule
\end{tabular}}
\label{tabdsd}
\end{table}\newpage

Note that, apart from three pairs of dimensions for which these
correlations are particularly high (1st and 2nd, 1st and 4th, and
3rd and 5th), the other correlations are smaller than 0.9. In
particular, the correlation between the 2nd and the 5th dimensions
is smaller than 0.3; moreover, the correlation is smaller than 0.7
for three other pairs of dimensions (1st and 5th, 2nd and 3rd, and
4th and 5th). This confirms that the dimensions found by the
clustering algorithm are actually distinct and then measuring the
health condition of elderly people necessarily requires a
multidimensional scale.
\section{Conclusions}
In the present paper we simultaneously study the issue of item
selection and of dimensionality of health status of elderly people
hosted in nursing homes. Our statistical approach is  based on the
joint use of latent class (LC) and item response theory (IRT)
models.

The study is based on a dataset collected in Italy within the
``Ulisse" project, which relies on a sample of 1699 elderly people
hosted in 37 nursing homes. The health status of these patients is
assessed by a set of items which are administered at repeated
occasions. From the original dataset, we extract 89 items, which
characterize different areas of the health status at the first
visit. In particular, we consider eight groups of items (each
corresponding to a different dimension), which measure: ({\em i})
cognitive conditions, ({\em ii}) auditory and view fields, ({\em
iii}) humor and behavioral disorders, ({\em iv}) activities of daily
living, ({\em v}) incontinence, ({\em vi}) nutritional field, ({\em
vii}) dental disorders, and ({\em viii}) skin conditions.

The analysis initially exploits the LC model for selecting a subset
of items which provides an amount of information close to that of
the full set of items. In particular, through the Bayesian
Information Criterion we find evidence of the presence of six latent
classes. Then, through the estimated conditional probabilities of
the LC model we start ranking the items according to their
discriminant power. This is measured by the standardized difference
between the estimated conditional probabilities across latent
classes. However, this selection process appears too conservative.

The items selection analysis is then completed by using an IRT model
based on a multidimensional 2PL parametrization. In particular, the
applied strategy first selects a benchmark model which has a number
of dimensions equal to the eight initial dimensions defined by the
questionnaire and then, by applying the 2PL model, selects the items
depending on their discriminatory power in measuring the latent
trait. The subset of selected items is then used to study the
dimensionality of the health condition of elderly people. A this
aim, we apply a hierarchical clustering algorithm which, starting
from the multidimensional model with eight dimensions, ends with the
unidimensional model. Within these two extremes we find all the
possible numbers of dimensions of the analyzed phenomenon and select
the most suitable by a series of likelihood ratio tests. On the
basis of this procedure we find evidence of five dimensions obtained
by collapsing the initial eight dimensions. These groups have the
following structure: $\{1\},\{2\},\{4,5\},\{3,8\},\{6,7\}$. These
five dimensions may be referred to as: ({\em i}) cognitive
conditions, ({\em ii}) auditory and view fields, ({\em iii})
activities of daily living and incontinence, ({\em iv}) humor and
behavioral disorders and skin conditions, and ({\em v}) nutritional
field and dental disorders.

Finally, the applied methodology suggests that the dimensionality of
health status of elderly people is a relevant aspect to be
considered in order to obtain a clear classification of the nursing
home facilities in the Italian context. Moreover, the IRT analysis
shows that the identified dimensions have not the same
discriminating power in determining the health status.

\bibliography{welfare}

\begin{thebibliography}{}

\bibitem[Alecxih and Kennell, 1994]{Alecxih1994}
Alecxih, L. and Kennell, D. (1994).
\newblock The economic impact of long-term care on individuals.
\newblock Technical report, US Department of Health and Human Services.

\bibitem[Anderson and Hussey, 2000]{Anderson2000}
Anderson, G.~F. and Hussey, P.~S. (2000).
\newblock Population aging: a comparison among industrialized countries.
\newblock {\em Health Aff (Millwood)}, 19(3):191--203.

\bibitem[Bartolucci, 2007]{barolucci2007}
Bartolucci, F. (2007).
\newblock A class of multidimensional irt models for testing unidimensionality
  and clustering items.
\newblock {\em Psychometrika}, 72(2):141--157.

\bibitem[Dempster et~al., 1977]{Dempster1977}
Dempster, A.~P., Laird, N.~M., and Rubin, D.~B. (1977).
\newblock Maximum likelihood from incomplete data via the em algorithm.
\newblock {\em Journal of the Royal Statistical Society. Series B
  (Methodological)}, 39(1):1--38.

\bibitem[Goodman, 1974]{goodman1974}
Goodman, L.~A. (1974).
\newblock Exploratory latent structure analysis using both identifiable and
  unidentifiable models.
\newblock {\em Biometrika}, 61(2):215--231.

\bibitem[Goodman, 1978]{goodman1978}
Goodman, L.~A. (1978).
\newblock {\em Analyzing qualitative/categorical data log-linear models and
  latent-structure analysis}.
\newblock Addison-Wesley Pub. Co.

\bibitem[Hambleton, 1996]{Hambleton1996}
Hambleton, R.~K., editor (1996).
\newblock {\em Handbook of Modern Item Response Theory}.
\newblock Springer, 1 edition.

\bibitem[Harrington et~al., 2003]{harrington2003}
Harrington, C., Collier, E., O'Meara, J., Kitchener, M., Simon, L.~P., and
  Schnelle, J.~F. (2003).
\newblock Federal and state nursing facility websites: just what the consumer
  needs?
\newblock {\em Am J Med Qual}, 18(1):21--37.

\bibitem[Kane, 1981]{kane1981}
Kane, R.~A. (1981).
\newblock Assuring quality of care and quality of life in long term care.
\newblock {\em QRB Qual Rev Bull}, 7(10):3--10.

\bibitem[Kane, 1998]{kane1998}
Kane, R.~L. (1998).
\newblock Assuring quality in nursing home care.
\newblock {\em J Am Geriatr Soc}, 46(2):232--237.

\bibitem[Lazarsfeld and Henry, 1968]{laza:henr:68}
Lazarsfeld, P.~F. and Henry, N.~W. (1968).
\newblock {\em Latent Structure Analysis}.
\newblock Houghton Mifflin, Boston.

\bibitem[Mor et~al., 2003]{VincentMor04012003}
Mor, V., Berg, K., Angelelli, J., Gifford, D., Morris, J., and Moore, T.
  (2003).
\newblock The quality of quality measurement in u.s. nursing homes.
\newblock {\em Gerontologist}, 43(90002):37--46.

\bibitem[Morris, 1997]{Morris:1997}
Morris, J.~N. (1997).
\newblock Development of the nursing home resident assessment instrument in the
  usa.
\newblock {\em Age and Agening}, pages 19--25.

\bibitem[Phillips et~al., 2007]{Phillips2007}
Phillips, C.~D., Hawes, C., Lieberman, T., and Koren, M.~J. (2007).
\newblock Where should momma go? current nursing home performance measurement
  strategies and a less ambitious approach.
\newblock {\em BMC Health Serv Res}, 7:93.

\bibitem[Schwarz, 1978]{schwarz1978}
Schwarz, G. (1978).
\newblock Estimating the dimension of a model.
\newblock {\em The Annals of Statistics}, 6(2):461--464.

\end{thebibliography}
\bibliographystyle{apalike}

\begin{table*}[h]
\centering
Appendix 1: Description of the selected Items\\
\tiny{
\begin{tabular}{r  r r l }
\\
\toprule
n item  &   Our &   Ulisse& item description                                \\
    &    class      & class     &                                           \\
        \hline
01  &   CC1     &   e2a &       recalls what recently happened (5 minutes)      \\
02  &   CC2     &   e2b &       keeps some past memories green              \\
03  &   CC3     &   e3a &       recalls the actual season                   \\
04  &   CC4     &   e3b &       recalls where is him room                   \\
05  &   CC5     &   e3c &       recalls the names and faces of the staff                \\
06  &   CC6     &   e3d &       recalls where he is                     \\
07  &   CC7     &   e4  &       decides about his daily  activities         \\
08  &   CC8     &   e5a &       gets  easily sidetracked        \\
09  &   CC9     &   e5b &       shows episodes of  altered
perception  or awareness of surrounding \\
10  &   CC10    &   e5c &       shows episodes of disorganized speech        \\
11  &   CC11    &   e5d &       has periods of restlessness movements       \\
12  &   CC12    &   e5e &       shows lethargic spans       \\
13  &   CC13    &   e5f &       does its cognitive conditions change during the day     \\
14  &   CAS1    &   f1  &       shows hearing deficiency        \\
15  &   CAS2    &   f4  &       making self-understood    \\
16  &   CAS3    &   f5  &       has a clear language        \\
17  &   CAS4    &   f6  &       is capable of understand others     \\
18  &   CAS5    &   g1  &       is able to see in conditions of adequate lighting       \\
19  &   HBD1    &   h1a &       made negative statements       \\
20  &   HBD2    &   h1b &       made repetitive questions        \\
21  &   HBD3    &   h1c &       made repetitive verbalizations        \\
22  &   HBD4    &   h1d &       shows persistent anger with self or others      \\
23  &   HBD5    &   h1e &       shows self deprecation disesteem       \\
24  &   HBD6    &   h1f &       expresses fears are not real        \\
25  &   HBD7    &   h1g &       believes itself to be dying     \\
26  &   HBD8    &   h1h &       complains about his health      \\
27  &   HBD9    &   h1i &       unpleasant mood in morning         \\
28  &   HBD10   &   h1j &       has problems with sleep     \\
29  &   HBD11   &   h1k &       suffers from insomnia       \\
30  &   HBD12   &   h1l &       has expressions of sad-faced        \\
31  &   HBD13   &   h1m &       easily tears        \\
32  &   HBD14   &   h1n &       shows repetitive movements     \\
33  &   HBD15   &   h1o &       abstains from activities of interest        \\
34  &   HBD16   &   h1p &       shows reduced local interactions        \\
35  &   HBD17   &   h4a1    &       wanders aimlessly       \\
36  &   HBD18   &   h4b1    &       uses offensive language     \\
37  &   HBD19   &   h4c1    &       is physically aggressive        \\
38  &   HBD20   &   h4d1    &       has a socially inappropriate behavior       \\
39  &   HBD21   &   h4e1    &       refuses assistance      \\
40  &   ADL1    &   j1a1    &       needs support in moving to/from lying position      \\
41  &   ADL2    &   j1b1    &       needs support in moving to/from bed, chair, wheelchair      \\
42  &   ADL3    &   j1c1    &       walks between different points within the room      \\
43  &   ADL4    &   j1d1    &       walks in the corridor       \\
44  &   ADL5    &   j1e1    &       walks into the nursing home ward        \\
45  &   ADL6    &   j1f1    &       walks outside the nursing home ward     \\
46  &   ADL7    &   j1g1    &       needs support for dressing      \\
47  &   ADL8    &   j1h1    &       needs support for eating        \\
48  &   ADL9    &   j1i1    &       needs support  using the toilet room        \\
49  &   ADL10   &   j1j1    &       needs support for personal hygiene      \\
50  &   ADL11   &   j2a1    &       needs support for taking full-body bath/shower      \\
51  &   ADL12   &   j3a1    &       shows balance problems      \\
52  &   ADL13   &   j4a1    &       shows loss of mobility in the neck      \\
53  &   ADL14   &   j4b1    &       shows loss of mobility in the arm including shoulder or elbow       \\
54  &   ADL15   &   j4c1    &       shows limitations in the movements of the hand including wrist or finger  \\
55  &   ADL16   &   j4d1    &       shows loss of mobility of the leg and hip       \\
56  &   ADL17   &   j4e1    &       shows loss of mobility of the foot and ankle        \\
57  &   ADL18   &   j4f1    &       shows limitations in other movements        \\
58  &   I1      &   k1a &       fecal incontinence      \\
59  &   I2      &   k1b &       urinary incontinence        \\
60  &   I3      &   k2a &       elimination of faeces (constipation)        \\
61  &   I4      &   k2b &       stipsi  \\
62  &   I5      &   k2c &       diarrhea        \\
63  &   I6      &   k2d &       fecaloma        \\
64  &   I7      &   k3b &       need aids (bladder retraining)      \\
65  &   I8      &   k3c &       need aids (external catheter)       \\
66  &   I9      &   k3d &       need aids (indwelling catheter)     \\
67  &   I10     &   k3e &       need aids (intermittent catheter)       \\
68  &   I11     &   k3f &       did not use toilet room/commode/urina    \\
69  &   I12     &   k3g &       pads/briefs used        \\
70  &   I13     &   k3h &       enemos/irrigations used    \\
71  &   I14     &   k3i &       ostomy present    \\
72  &   I15     &   k3j &       need aids (others)      \\
73  &   N1      &   n1a &       chewing problem        \\
74  &   N2      &   n1b &       swallowing problem        \\
75  &   N3      &   n1c &       mouth pain    \\
76  &   N4      &   n3a &       lose weight     \\
77  &   N5      &   n3b &       gain weight     \\
78  &   N6      &   n4a &       complain about the taste of many foods      \\
79  &   N7      &   n4b &       complains of being hungry       \\
80  &   N8      &   n4c &       leave the food on his plate     \\
81  &   D1      &   o1a &       debris (soft, easily movable substances) present in mouth prior to going to bed at night        \\
82  &   D2      &   o1b &       has dentures or removable bridge        \\
83  &   D3      &   o1c &       some/all natural teeth lost—does not have or does not use dentures (or partial plates)  \\
84  &   D4      &   o1d &       broken, loose, or carious teeth        \\
85  &   D5      &   o1e &       Inflamed gums (gingiva); swollen or bleeding gums; oral abscesses; ulcers or rashes \\
86  &   D6      &   o1f &       daily cleaning of teeth/dentures or daily mouth care—by resident or staff       \\
87  &   SK1     &   p2a &       pressure ulcer    \\
88  &   SK2     &   p2b &       stasis ulcers       \\
89  &   SK3     &   p3  &       had an ulcer that was resolved or cured
    \\
\bottomrule
\end{tabular}}
\end{table*}

\end{document}